\documentclass{article}

\usepackage{arxiv}

\usepackage[utf8]{inputenc} 
\usepackage[T1]{fontenc}    
\usepackage{hyperref}       
\usepackage{url}            
\usepackage{booktabs}       
\usepackage{amsfonts}       
\usepackage{nicefrac}       
\usepackage{microtype}      
\usepackage{lipsum}
\usepackage{graphicx}
\usepackage{indentfirst}
\usepackage{multirow}
\usepackage[square,sort,comma,numbers]{natbib}
\usepackage{siunitx}  
\usepackage{amssymb}  
\usepackage{amsmath}
\usepackage[title]{appendix}%
\title{3D-Mol: A Novel Contrastive Learning Framework for Molecular Property Prediction with 3D Information}

\author{
  Taojie Kuang \\
  Peng Cheng Laboratory\\
  South China University of Technology\\
  \texttt{kuangtj@pcl.ac.cn} \\
  \And
  Yiming Ren \\
  Peng Cheng Laboratory\\
  \texttt{renym@pcl.ac.cn} \\
  \And
  Zhixiang Ren\thanks{Corresponding author} \\
  Peng Cheng Laboratory\\
  \texttt{renzhx@pcl.ac.cn} \\
}

\begin{document}
\maketitle

\begin{abstract}
Molecular property prediction, crucial for early drug candidate screening and optimization, has seen advancements with deep learning-based methods. 
While deep learning-based methods have advanced considerably, they often fall short in fully leveraging 3D spatial information. 
Specifically, current molecular encoding techniques tend to inadequately extract spatial information, leading to ambiguous representations where a single one might represent multiple distinct molecules. 
Moreover, existing molecular modeling methods focus predominantly on the most stable 3D conformations, neglecting other viable conformations present in reality. 
To address these issues, we propose 3D-Mol, a novel approach designed for more accurate spatial structure representation. 
It deconstructs molecules into three hierarchical graphs to better extract geometric information. 
Additionally, 3D-Mol leverages contrastive learning for pretraining on 20 million unlabeled data, treating their conformations with identical topological structures as weighted positive pairs and contrasting ones as negatives, based on the similarity of their 3D conformation descriptors and fingerprints.
We compare 3D-Mol with various state-of-the-art baselines on 7 benchmarks and demonstrate our outstanding performance. 
\end{abstract}

\section{Introduction}

\begin{figure}[t!]
\centering
    \includegraphics[width=0.7\linewidth]{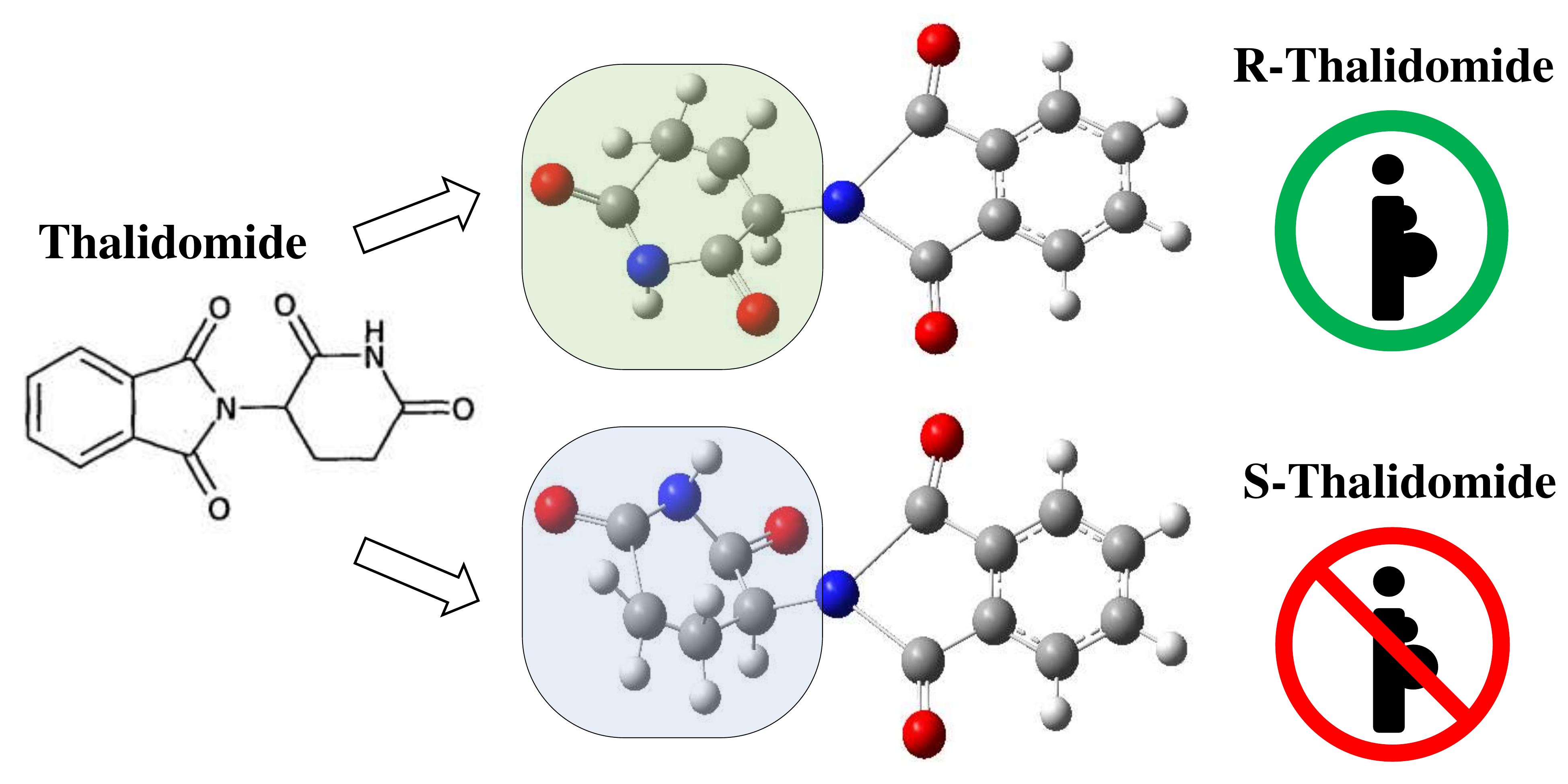}
    \caption{\textbf{Geometric difference leads to diverse properties.} 
    Thalidomide exists in two distinct 3D stereoisomeric forms, known as R-Thalidomide and S-Thalidomide. 
    These two molecules can be represented by the same SMILES, but they have significantly dissimilar properties.
    The former is recognized for its therapeutic properties, while the latter has been implicated in teratogenesis.
    }
    \label{fig:fig1}
\end{figure}

Molecular property prediction can effectively accelerate drug discovery by prioritizing promising compounds, streamlining drug development and increasing success rates. 
Moreover, it contributes to the comprehension of structure-activity relationships by demonstrating the influence of particular features on molecular interactions and other biological effects.
Recently, deep learning methods have significantly advanced molecular property prediction, providing enhanced accuracy and deeper insights into complex molecular behaviors. 
The integration of 3D molecular information, which includes a comprehensive view of molecular structures, significantly enhance the model's understanding of molecular properties and interactions.
However, the expensive and time-consuming experiments result in the scarcity of labeled data, which significantly constrains the capacity of deep learning methods to extract 3D spatial information.\\
\indent To fully understand the knowledge in unlabeled data, numerous methods based on self-supervised learning have been proposed to enhance the performance of molecular property prediction. 
For example, early work\citep{gohSMILES2VecInterpretableGeneralPurpose2017,huang2020deeppurpose,chithranandaChemBERTaLargeScaleSelfSupervised2020} employed self-supervised learning approaches for processing data represented in the Simplified Molecular Input Line Entry System (SMILES)\citep{weininger1988smiles}. 
However, SMILES is not adequate for the representation of the topological structure of molecule, making it challenging to provide reliable results. 
In parallel, various self-supervised learning methods based on molecular graph\citep{pmlr-v70-gilmer17a,yang2019analyzing,huStrategiesPretrainingGraph2019,liu2019n,xiong2019pushing,wangMolecularcontrastivelearning2022,NEURIPS2020_94aef384} despite employing molecular graph data to encode the topological structure of molecule, neglects the critical three-dimensional spatial information of the molecule. 
Since different 3D structures may lead to dissimilar molecular properties despite having the same 2D molecular topology. 
As an example shown in Figure \ref{fig:fig1}, Thalidomide, a sedative treatment for morning sickness in pregnant women in the 1950s, has two distinct 3D structures, R-Thalidomide and S-Thalidomide. 
The former has desired drug effects, while the latter has been implicated in teratogenesis.
Recently, several works\citep{schutt2017schnet,gasteiger2020directional,shui2020heterogeneous,danel2020spatial,gasteiger2020directional,fang2022geometry,zhou2023uni} utilizing molecular 3D structures have been introduced. However, limited by pretraining methods, they have not fully learned the 3D spatial information in unlabeled data.
Specifically, these methods focus only on the most stable (lowest energy) 3D conformations, neglecting other existing conformations.
Therefore, it is imperative to develop an approach that comprehensively acquires 3D analytical insights, encompassing both pretrain strategy and encoding technique.\\
\indent To address these issues, we propose a novel framework, 3D-Mol, for molecular representation and property prediction. 
We employ three graphs to hierarchically represent the atom-bond, bond-angle, and dihedral information of molecule, integrating information from these hierarchies through a message-passing strategy to obtain a comprehensive molecular representation. 
Moreover, by using a vast amount of unlabeled data, we create a novel self-supervised method, weighted contrastive learning, to pretrain our molecular encoder alongside the geometric approach from GearNet\citep{zhang2022protein}. 
In the proposed contrastive learning, conformations derived from the same SMILES are considered weighted positive pairs, while different ones are treated as weighted negative pairs, with weights indicating 3D conformation descriptor/fingerprint similarity.
The molecular encoder is then finetuned on downstream tasks to predict molecular properties. 
Finally, we compare our approach with several state-of-the-art(SOTA) baselines on 7 molecular property prediction benchmarks\citep{wu2018moleculenet}, where our method achieves the best results on 5 benchmarks. 
In summary, our main contributions are as follows:\\
\indent $\bullet$ We propose a novel molecular embedding method based on \textbf{hierarchical graph representation} to thoroughly extract the 3D spatial structural features of molecule.\\
\indent $\bullet$ We improve the contrastive learning approach by utilizing \textbf{3D conformational information} by considering conformations with the same SMILES as positive pairs and the opposites as negative pairs, while keeping the weight to indicate the 3D conformation descriptor and fingerprint similarity.\\
\indent $\bullet$ We evaluate 3D-Mol on various molecular property prediction benchmarks, showing that our model can significantly \textbf{outperform existing competitive models} on multiple tests.
\section{Related Work}
Due to the unique nature of molecular representation and the scarcity of labeled molecular data, existing methods generally use two methods to enhance the performance of molecular property prediction.
One insight entails the development of a novel molecular encoder tailored to molecular data for efficient molecular information extraction. 
The other emphasizes fully harnessing the potential of unlabeled data, typically by devising a unique pretraining approach to pretrain the molecular encoder using a large amount of unlabeled data. 
Details of the key components for each strategy are listed below.
\subsection{Molecular Representation and Encoder}
Molecular representation and encoding are essential for accurate property prediction, vital in applications like molecular design and drug discovery. 
Some early works\citep{cereto2015molecular,coley2017convolutional} learned representation from chemical fingerprints (FP), such as ECFP\citep{rogers2010extended} and MACCS\citep{durant2002reoptimization}. 
Other works learned representation from molecular descriptors, such as SMILES. Inspired by mature NLP models, SMILES-BERT\citep{wang2019smiles} used SMILES to extract molecular representations by applying the BERT\citep{devlinBERTPretrainingDeep2018} pretrain strategy. 
However, these methods depend on feature engineering, failing to capture the complete topological structure information of molecule.\\
\indent Since a molecular graph is a natural representation of a molecule and conveys topological information, several research in recent years have embraced it as a means of molecular representation. 
GG-NN\citep{pmlr-v70-gilmer17a}, DMPNN\citep{yang2019analyzing}, and DeepAtomicCharge\citep{wang2021deepatomiccharge} employed a message passing strategy for molecular property prediction. 
AttentiveFP\citep{xiong2019pushing} used a graph attention network to aggregate and update node information. 
The MP-GNN\citep{li2022multiphysical} merged specific-scale graph neural network (GNN) and element-specific GNN, capturing various atomic interactions of multiphysical representations at different scales. 
MGCN\citep{lu2019molecular} designed a graph convolution network to capture multilevel quantum interactions from the conformation and spatial information of molecule. \\
\indent The works mentioned above focus on 2D molecular representation, which might miss crucial chemical details\citep{qiao2020orbnet} and prove insufficient for accurate molecular property prediction\citep{li2022deep}.
Recently, some studies have attempted to enhance performance by modeling 3D molecular structure. 
Significant efforts have been made to employ 3D voxel-based representations for understanding molecular structures. 
Stepniewska et al. \citep{stepniewska2018development} used 3D convolutions to estimate the binding affinity of ligand-receptor complexes. 
libmolgrid \citep{sunseri2020libmolgrid} provided a library representing 3D molecular structures as multidimensional voxelized grids. 
OctSurf \citep{liu2021octsurf} used an octree-based representation to describe the interaction between protein pockets and ligands. 
In addition to voxel-based representations, several methods have been developed to embed 3D molecular information directly into GNNs.
SGCN\citep{danel2020spatial} applied different weights according to atomic distances during the GCN-based message passing process. 
SchNet\citep{schutt2017schnet} modeled complex atomic interactions using Gaussian radial basis functions for potential energy surface prediction to accelerate the exploration of chemical space. 
DimeNet\citep{gasteiger2020directional} proposed directional message passing to fully utilize directional information within molecule. 
GEM\citep{fang2022geometry} developed a novel geometrically enhanced molecular representation learning method and employs a specifically designed geometric-based GNN structure. 
However, these methods do not fully exploit the 3D structural information(like dihedral angle) of molecule. 
\subsection{Self-supervised Learning on Molecule}
\indent Self-supervised learning, with its substantial success in various research domains, has inspired numerous molecular property prediction studies. 
Influenced by methods such as BERT\citep{devlinBERTPretrainingDeep2018} and GPT\citep{floridi2020gpt}, these studies employ this approach to efficiently harness large volumes of unlabeled data for pretraining.
\indent For one-dimensional data, SMILES is frequently used to extract molecular representations in the pretraining stage. 
SMILES2Vec\citep{gohSMILES2VecInterpretableGeneralPurpose2017} employed the RNN to extract features from SMILES. 
ChemBERTa\citep{chithranandaChemBERTaLargeScaleSelfSupervised2020} followed RoBERTa\citep{liu2019roberta} by employing masked language modeling as a pretraining task, predicting masked tokens to restore the original sentence, which helped models understand sequence semantics. 
SMILES Transformer\citep{honda2019smiles} used a SMILES string as input to produce a temporary embedding, which is then restored to the original input by a decoder.\\
\indent As the topological information of molecular graphs received greater attention, numerous pretraining methods focused on molecular graph data have been proposed. 
N-gram graph\citep{liu2019n} used the n-gram method in NLP to extract representations of molecule. 
PretrainGNN\citep{huStrategiesPretrainingGraph2019} proposed a new pretrain strategy, including node-level and graph-level self-supervised pretraining tasks. 
GraphCL\citep{you2020graph}, MOCL\citep{sun2021mocl}, and MolCLR\citep{wangMolecularcontrastivelearning2022} performed molecular contrastive learning via GNN by proposing new molecular graph augmentation methods. 
MPG\citep{li2021effective} and GROVER\citep{NEURIPS2020_94aef384} focused on node level and graph level representation and designed corresponding pretraining tasks at both levels.
iMolCLR\citep{wang2022improving}, Sugar\citep{sun2021sugar} and ReLMole\citep{ji2022relmole} focused on the substructure of molecule, and designed the substructure pretraining task by using substructure information. \\
\indent With the 3D structure information of molecule proven to boost molecular property prediction, recent works have focused on pretraining tasks for the 3D structure information of molecule. 
3DGCN\citep{cho2019enhanced} introduced a relative position matrix that includes 3D positions between atoms to ensure translational invariance during convolution. 
GraphMVP\citep{liu2021pre} proposed the SSL method involving contrastive learning and generative learning between 3D and 2D molecular views. 
GEM\citep{fang2022geometry} proposed a self-supervised framework using molecular geometric information by constructing a new bond angle graph, where the chemical bonds within a molecule are considered as nodes and the angle formed between two bonds is considered as the edge. 
Uni-Mol\citep{zhou2023uni} employed a transformer model to extract molecular representation by predicting atom distance. 
However, These works utilize only the most stable 3D conformation, thereby overlooking other conformations that exist in the real world.
\section{Method}
\begin{figure}[!h]
    \centering
    \includegraphics[width=1.0\linewidth]{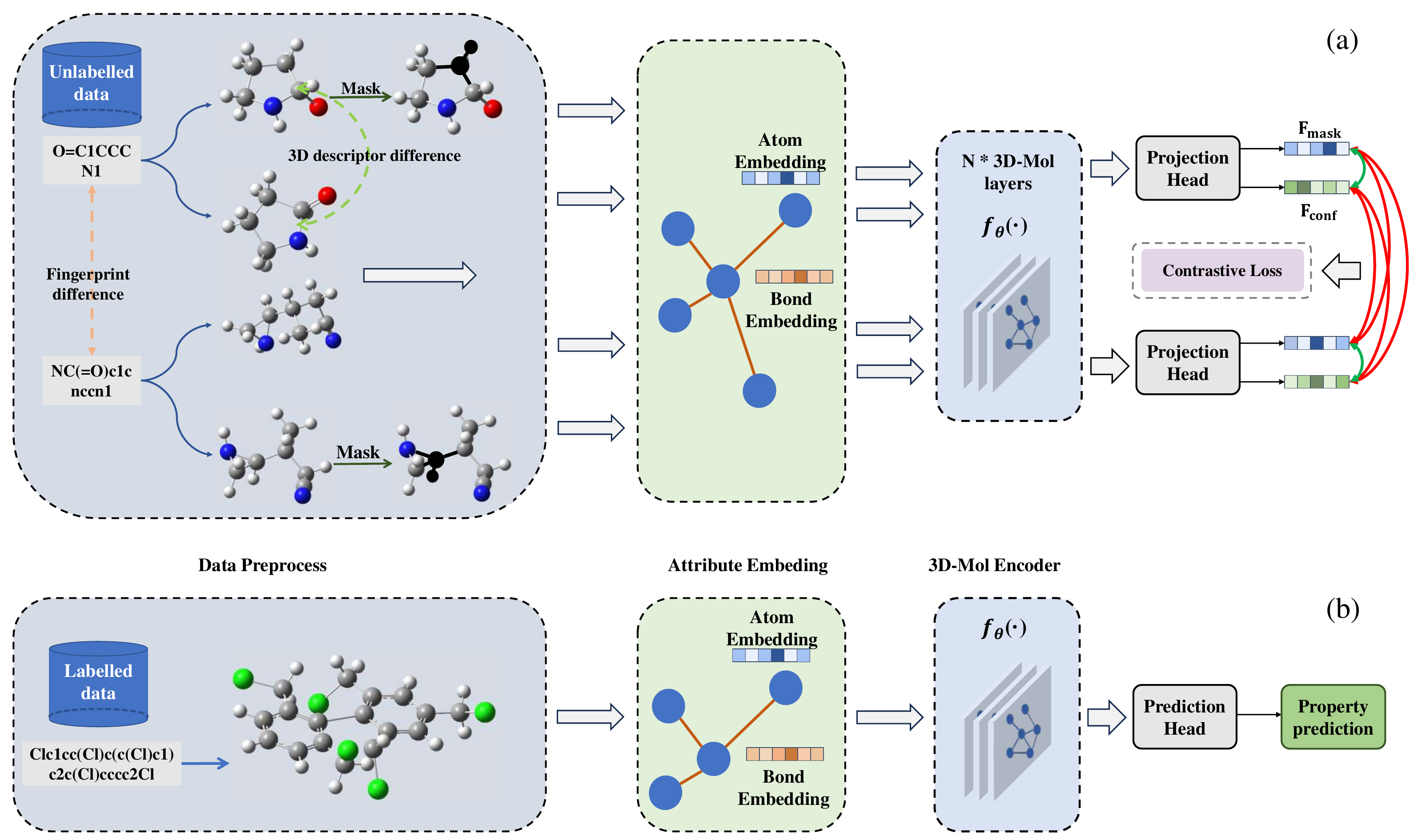}
    \caption{\textbf{The overview of the 3D-Mol model framework.} 
    a) In the pretraining stage, we employ weighted contrastive learning to effectively pretrain our model. 
    In addition to using the mask strategy for graph data augmentation, we consider conformations from the same SMILES as positive pairs, while the weight represents their 3D conformation descriptor similarity. 
    Conversely, distinct topological structures are treated as negative pairs, and the weight is dependent on fingerprint differences. 
    b) In the finetuning stage, we use one well-pretrained encoder model to refine our approach across diverse downstream datasets through supervised learning.}
    \label{fig:fig2}
\end{figure}

This section outlines the creation of 3D-Mol, a framework designed for 3D structural molecular property prediction, focusing on hierarchical graph-based molecular representation and the strategic weighting of contrastive pairs. Figure \ref{fig:fig2} provides an overview, with subsequent parts delving into specifics.
\subsection{Molecular Encoder}
\subsubsection{Hierarchical Graph}
\indent Molecular raw data is represented by SMILES in most molecular databases. 
To extract spatial structure information from molecule, we use RDKit\citep{landrum2013rdkit} to transform the SMILES representation into 3D molecular conformations. 
To fully extract 3D construct information, we deconstruct molecular conformation into three hierarchical graphs, denoted as $Mol= \{G_{a-b}, G_{b-a}, G_{d-a}\}$. 
The atom-bond graph, commonly used as a 2D molecular graph, is represented as $G_{a-b}=\{V, E, P_{atom}, P_{bond}\}$, where $V$ is the set of atoms and $E$ is the set of bonds. 
$P_{atom} \in R^{|V|*d_{atom}}$ are the attributes of atoms, and $d_{atom}$ is the number of atom attributes. 
$P_{bond} \in R^{|E|*d_{bond}}$ are the attributes of bonds, and $d_{bond}$ is the number of bond attributes. 
The bond-angle graph, is represented as $G_{b-a}=\{E, P, Ang_{\theta}\}$, 
where $P$ is a set of the plane that is comprised of 3 connected atoms. 
$Ang_{\theta}$ is the set of corresponding bond angles $\theta$. 
The dihedral-angle graph, is represented as $G_{d-a}=\{P, D, Ang_{\phi}\}$. 
The attributes of the plane are the attributes of 3 connected atoms and the corresponding bonds.
$D$ represents the set of two connected planes, which connect with a bond. 
$Ang_{\phi}$ represents the corresponding dihedral angle $\phi$. 
These three graphs represent an actual molecule, and help our encoder learn 3D structure information.
\subsubsection{Attribute Embedding}
\indent The 3D information of the molecule, such as the length of bonds and the angle between bonds, carries key chemical information. 
Firstly, we convert spatial characteristics to latent vectors. 
Referring to the previous work\citep{shui2020heterogeneous}, we employed RBF(Radial basis function) layers to encode different geometric factors:
\begin{equation}
    F^k_{l} = exp(-\beta^{k}_{l}(exp(-l)-\mu^{k}_{l})^2) * W^k_{l}
\end{equation}
where $F^k_{l}$ is the k-dimensional feature of bond length $l$, and $\mu^{k}_{l}$ and $\beta^{k}_{l}$ are the center and width of $l$ respectively. 
$\mu^{k}_{l}$ is 0.1$k$ and $\beta^{k}_{l}$ is 10. 
Similarly, the k-dimensional features of $F^k_{\theta}$ and $F^k_{\phi}$ of x are computed as:
\begin{equation}
    F^k_{\theta} = exp(-\beta^{k}_{\theta}(-\theta-\mu^{k}_{\theta})^2) * W^k_{\theta}
\end{equation}
\begin{equation}
    F^k_{\phi} = exp(\beta^{k}_{\phi}(-\phi-\mu^{k}_{\phi})^2) * W^k_{\phi}
\end{equation}
Where $\mu^{k}_{\theta}$ and $\mu^{k}_{\phi}$ are denoted as the centers of bond angles and dihedral angles, respectively, establishing the peak of the function and centralizing the feature transformation. Similarly, the widths that determine the spread of the RBF, are represented as $\beta^{k}_{\theta}$ for bond angles and $\beta^{k}_{\phi}$ for dihedral angles. These widths dictate the spread of the function. The numerical values for these centers are set at $\pi$/K, where K is the number of feature dimensions.\\
\begin{figure}[]
    \centering
    \includegraphics[width=1.0\linewidth]{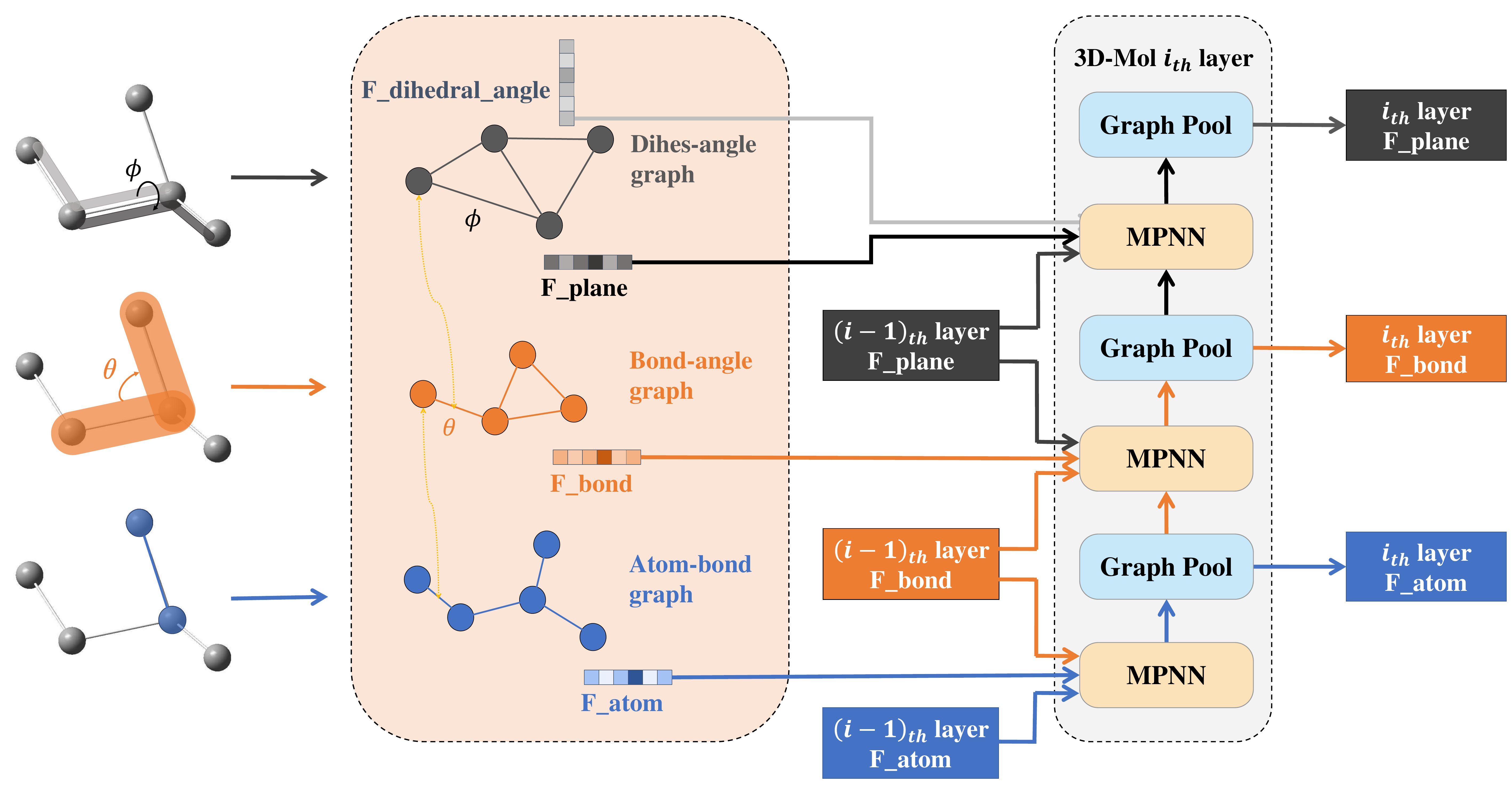}
    \caption{\textbf{The overview of the 3D-Mol encoder layer.} The 3D-Mol encoder layer comprises three steps. Firstly, employing a message passing strategy, nodes in each graph exchange messages with their connected edges, leading to the updating of edge and node latent vectors. Secondly, the edge latent vector from the lower-level graph is transmitted to the higher-level graph as part of the node latent vector. Finally, the iteration is performed n times to derive the $n_{th}$ node latent vector, from which we extract the molecular latent vectors.}
    \label{fig:fig3}
\end{figure}
\indent For the other attributes of atom and bond, we represent them with $P_{atom}$ and $P_{bond}$ and embed them with the word embedding function. 
The initial features of atoms and bonds are represented as $F^0_{atom}$ and $F^0_{bond}$ respectively.
\subsubsection{Graph Embedding}
\indent To embed the molecular hierarchical graph, we employ message passing strategy in $\{G^i_{a-b}, G^i_{b-a}, G^i_{d-a}\}$. 
For the $i_{th}$ layer in 3D-Mol, the information of those graphs will be updated by graph neural network. 
The overview is shown in figure \ref{fig:fig3}, and the details are as follows:\\
\indent First, we use $GNN^i_{a-b}$ to aggregate the atom and bond latent vectors in $G^i_{a-b}$. 
Given an atom v, its representation vector $F^{i}_{v}$ is formalized by:
\begin{equation}
    a^{i, a-b}_{v} = Agg^{(i)}_{a-b}({F^{i-1}_{v}, F^{i-1}_{u}, F^{i-1}_{uv}|u \in N(v)})
\end{equation}
\begin{equation}
    F^{i}_{v} = Comb^{(k)}_{a-b, n}(F^{i-1}_{v}, a^{i}_{v})
\end{equation}
\begin{equation}
    F^{i, temp}_{uv} = Comb^{(k)}_{a-b, e}(F^{i-1}_{uv}, F^{i-1}_{u}, F^{i-1}_{v})
\end{equation}
where $N(v)$ is the set of neighbors of atom v in $G^i_{a-b}$, and $Agg^{(i)}_{a-b}$ is the aggregation function for aggregating messages from the atom neighborhood. 
$Comb^{(k)}_{a-b, n}$ and $Comb^{(k)}_{a-b, e}$ are the update functions for updating the latent vectors of atom and bond, respectively. 
$a^{i, a-b}_{v}$ is the information from the neighboring atom and the corresponding bond after being aggregated. 
$F^{i, temp}_{uv}$ is the temporary bond latent vectors of bond $uv$ in $i_{th}$ layer and is part of the bond latent vectors in $G^i_{b-a}$. \\
\indent Then, we use $GNN^i_{b-a}$ to aggregate the bond and plane vectors in $G^i_{b-a}$. Given a bond $uv$, its latent vector $F^{i}_{uv}$ is formalized by:
\begin{equation}
    \begin{split}
        a^{i, b-a}_{uv} = Agg^{(i)}_{b-a}(&\{F^{i-1}_{uv}, F^{i-1}_{vw}, F^{i-1}_{uvw}|u \in N(v) \cap \\
          & w \in N(v) \cap u \neq w\})
    \end{split}
\end{equation}
\begin{equation}
        F^{i}_{uv} = Comb^{(k)}_{b-a, n}(F^{i-1}_{uv}, F^{i, temp}_{uv}, a^{i}_{uv})
\end{equation}
\begin{equation}
    F^{i-1, temp}_{uvw} = Comb^{(k)}_{b-a, e}(F^{i-1}_{uvw}, F^{i-1}_{uv}, F^{i-1}_{vw})
\end{equation}
where $Agg^{(i)}_{b-a}$ is the aggregation function for aggregating messages from the bond neighborhood.  $Comb^{(k)}_{b-a, n}$ and $Comb^{(k)}_{b-a, e}$ are the update functions for updating the bond and plane latent vectors. 
$a^{i, b-a}_{uv}$ is the information from the neighboring bond and the corresponding bond angle after being aggregated. 
$F^{i-1, temp}_{uvw}$ is the temporary plane latent vectors of plane $uvw$ in $i_{th}$ layer and is part of the plane latent vectors in $G^i_{d-a}$. \\
\indent After processing the $G^i_{b-a}$, we use $GNN^i_{d-a}$ to aggregate the plane latent vector in $G^i_{d-a}$. 
Given a plane constructed by nodes u, v, w and bonds $uv$, $vw$, its latent vector $F^{i}_{uvw}$ is formalized by:
\begin{equation}
    \begin{split}
        a^{i, d-a}_{uvw} = Agg^{(i)}_{d-a}(\{F^{i-1}_{uvw}, F^{i-1}_{vwh}, F^{i-1}_{uvwh}|u \in N(v)  \cap \\
        v \in N(w) \cap w \in N(h) \cap u \neq v \neq w \neq h\}) 
    \end{split}
\end{equation}
\begin{equation}
    F^{i}_{uvw} = Comb^{(k)}_{d-a, n}(F^{i-1}_{uvw}, F^{i, temp}_{uvw}, a^{i}_{uvw})
\end{equation}
where $agg^{(i)}_{d-a}$ is the aggregation function for aggregating messages from the plane neighborhood. 
$Comb^{(k)}_{d-a, n}$ is the update functions for updating the plane latent vector. 
$a^{i}_{uvw}$ is the information from the neighboring plane and the corresponding dihedral angle after being aggregated. \\
\indent The representation vectors of the atoms at the final iteration are integrated to gain the molecular graph representation vector $F_{mol}$ by the Readout function, which is formalized as:
\begin{equation}
    F_{mol} = Readout({F^{n}_{u}|u \in V})
\end{equation}
where $F^{n}$ is the last 3D-Mol layer output. The molecular latent vector $F_{mol}$ is used to predict molecular properties.
\subsection{Pretrain Strategy}
\begin{figure}[]
    \centering
    \includegraphics[width=0.75\linewidth]{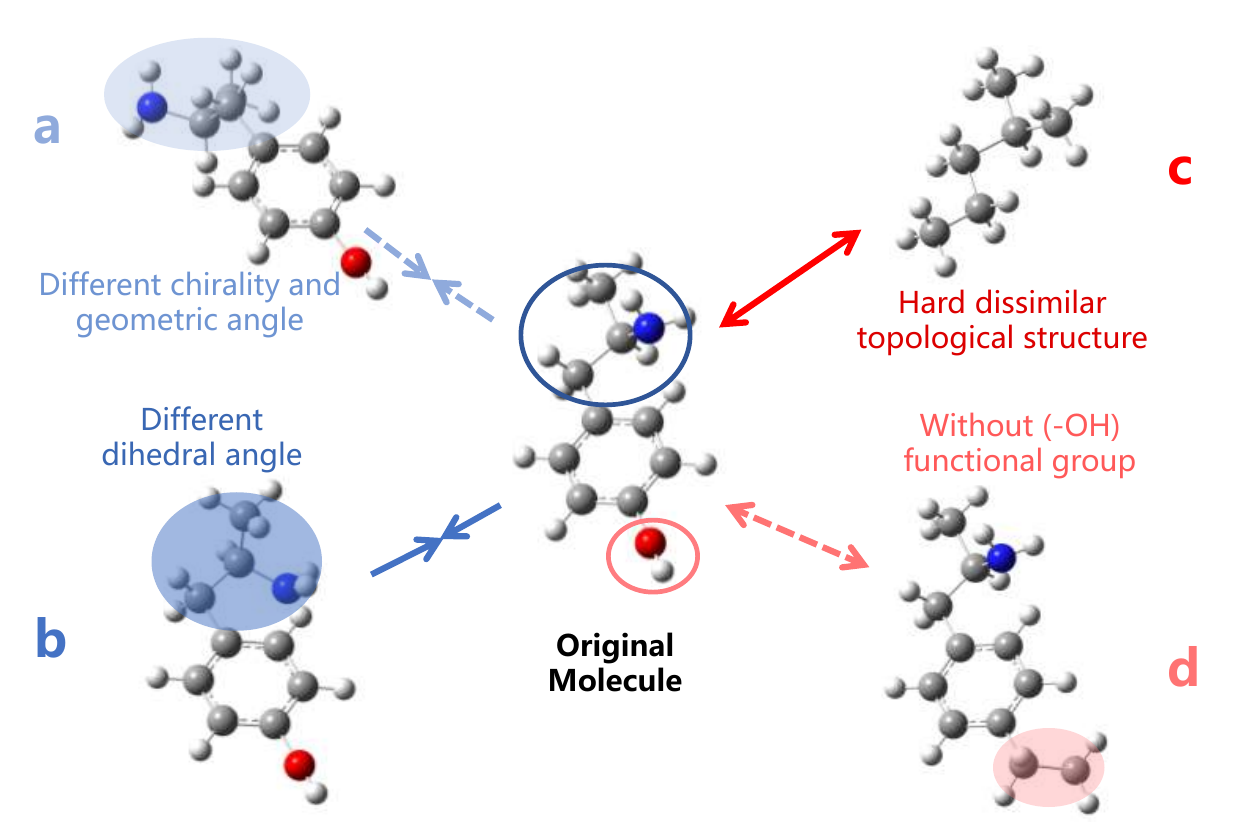}
    \caption{\textbf{The overview of weighted positive/negative pairs.} Using the molecule Oc1ccc(cc1)CC[NH3+]. \textbf{a). Low weight positive pairs:}  Features two conformations with the same SMILES and significant structural differences (e.g., chirality, geometric angles). \textbf{b). High weight positive pairs:} Depicts conformations with the same SMILES but minor differences, such as slight dihedral angle variations. \textbf{c). Low weight negative pair:} Shows conformations from different SMILES with similar scaffolds but missing little functional group, like an (-OH) group. \textbf{d). High weight negative pair:} Illustrates conformations from different SMILES, differing greatly in both scaffold and functional groups.}
    \label{fig:fig4}
\end{figure}

\indent To improve the performance of the 3D-Mol encoder, we employ contrastive learning for pretraining, categorizing conformations with identical topological structures as weighted positive pairs and contrasting ones as negatives, as shown in figure \ref{fig:fig4}.
Inspired by GearNet\citep{zhang2022protein}, we also combine our pretraining method with self-supervised tasks based on physicochemical and geometric properties. \\
\indent Our objective is to facilitate the learning of the consistency and differences between molecular 3D conformations. To accomplish this, we employ weighted contrastive learning using a batch of molecular representations, with the loss function defined as follows:
\begin{equation}
    L_{i, j}^{conf} = - log \frac{exp(w^{conf}_{i, j}sim(F_i, F^{mk}_j)/\tau)}{\Sigma_{k=1}^{2N}1{\{k \neq i\}}exp(w^{fp}_{i, k}sim(F_i, F_k)/\tau)}
\end{equation}
\begin{equation}
    w^{conf}_{i, j} = \lambda_{conf} * Sim_{dsp}(Dsp_i, Dsp_j)
\end{equation}
\begin{equation}
    w^{fp}_{i, k} = 1 - \lambda_{fp} * Sim_{FP}(Mconf_i, Mconf_k)
\end{equation}
where the two conformations with same SMILES, denoted as $Mconf_i$ and $Mconf_j$. $F_i$ is the latent vector extracted from $Mconf_i$, and $sim()$ measures the similarity between latent vectors, penalized by a weight coefficient $w^{conf}_{i, j}$, which is computed based on the 3D conformation descriptor similarity between $Mconf_i$ and $Mconf_j$. 
$w^{conf}_{i, j}$ represents the similarity between $Dsp_i$ and $Dsp_j$, which correspond to the 3D conformation descriptors of $Mconf_i$ and $Mconf_j$.
$Sim_{dsp}()$ evaluates the similarity between 3D conformation descriptors, and $\lambda_{conf} \in [0, 1]$ is the hyperparameter that determines the scale of penalty for the similarity between two conformations. 
$\tau$ is the temperature parameter.
In addition to using different conformations as the positive pair, we also employ node masking as a molecular data augmentation strategy. 
We random select 15$\%$ of atoms, mask them and their corresponding bonds, and the masked molecular $Mconf_j$ latent vector is denoted as $F^{mk}_j$.
The similarity measurement between two latent vectors $F_i$, $F_k$ from a negative molecule pair ($Mconf_i, Mconf_k$) is penalized by a weight coefficient $w^{fp}_{i, k}$, which computed by molecular fingerprint similarity between $Mconf_i$ and $Mconf_k$. 
$Sim_{FP}()$ evaluates similarity between molecular fingerprints, and $\lambda_{fp} \in [0, 1]$ is the hyperparameter that determines the scale of penalty for faulty negatives.
The details of the 3D conformation descriptors and fingerprint are shown in Appendix A.\\
\indent Since physicochemical and geometric information has been demonstrated to be important for molecular property prediction, we also employ geometry tasks as the pretraining method. 
For bond angle and dihedral angle prediction, we sample adjacent atoms to better capture local structural information. 
Since angular values are more sensitive to errors in protein structures than distances, we use discretized values for prediction. The following are the loss functions for the local geometry task:
\begin{equation}
    L_{i, j}^{l} = (f_{l}(Fn^{mk}_{n, i}, Fn^{mk}_{n, j}) - l_{i, j})^2
\end{equation}
\begin{equation}
    L_{i, j, k}^{\theta} = CE(f_{\theta}(Fn^{mk}_{n, i}, Fn^{mk}_{n, j}, Fn^{mk}_{n, k}), bin({\theta}_{i, j, k}))
\end{equation}
\begin{equation}
    \begin{split}
        L_{i, j, k, p}^{\phi} = CE(&f_{\phi}(Fn^{mk}_{n, i}, Fn^{mk}_{n, j}, Fn^{mk}_{n, k}, Fn^{mk}_{n, p}),\\& bin({\phi}_{i, j, k, p}))
    \end{split}
\end{equation}
where $f_{\phi}()$, $f_{\theta}()$ and $f_{l}$ are the MLPs for the local geometry task, and $L_{i, j}^{l}$, $L_{i, j, k}^{\theta}$, $L_{i, j, k, p}^{\phi}$ and $L_i^{FP}$ are the loss functions for each task. $CE()$ is the cross entropy loss, and $bin()$ is used to discretize the bond angle and dihedral angle. $Fn^{mk}_{n, i}$ is the latent vector of node i after masking the corresponding sampled items in each task.\\
\indent In addition to the aforementioned pretraining tasks to capture global molecular information, we leverage masked molecular latent vectors for FP prediction and atom distance prediction, effectively incorporating latent representations to enrich the predictive capability. The following are the loss functions for the global geometry task:
\begin{equation}
    L_i^{FP} = BCE(f_{FP}(F^{mk}), FP_i)
\end{equation}
\begin{equation}
    L_{i,j}^{dist} = (f_{dist}(Fn^{mk}_{n, i}, Fn^{mk}_{n, j}) - dist_{i, j})^2
\end{equation}
where $f_{FP}$ and $f_{dist}()$ are the MLPs for global geometric tasks, and $L_i^{FP}$ and $L_{i,j}^{dist}$ are the loss functions for each task. $BCE()$ is binary cross entropy loss. $F^{mk}$ is the latent vector of the masking molecule.\\
\indent In the culmination of our pretraining stage, we consolidate the various loss functions into a unified training objective through an uncertainty weighted sum approach. The following is the final loss function:
\begin{equation}
    \begin{split}
    L^{final} = &L^{FP}/{\sigma_{FP}^2} + L^{dist}/{\sigma_{dist}^2} + L^{\phi}/{\sigma_{\phi}^2} + L^{\theta}/{\sigma_{\theta}^2} + L^{l}/{\sigma_{l}^2} + L^{conf}/{\sigma_{conf}^2}\\& + log\sigma_{FP} + log\sigma_{dist} + log\sigma_{\phi} + log\sigma_{\theta} + log\sigma_{l} + log\sigma_{conf}
    \end{split}
\end{equation}
Where $L^{final}$ is the final loss we use to train the encoder in pretraining stage, and the $\sigma_{FP}$, $\sigma_{dist}$, $\sigma_{\phi}$, $\sigma_{\theta}$, $\sigma_{l}$ and $\sigma_{conf}$ are the uncertainty associated with each loss component, representing the model's confidence in each of these loss terms.
This method employs individual uncertainty terms for each loss component, allowing the model to dynamically adjust the influence of each based on its confidence in the respective predictions, which facilitates a balanced optimization across diverse molecular features, from spatial arrangements to angular orientations.
\section{Experiment}
\label{sec:eva}
\indent 
In this section, we conduct experiments on 7 benchmark datasets in MoleculeNet\citep{wu2018moleculenet} to demonstrate the effectiveness of our method for molecular property prediction. 
We use a large amount of unlabeled data and our pretrain strategy to pretrain our encoder, then use the downstream task to finetune the well-pretrained model and predict the molecular property. 
We compare it with a variety of SOTA methods and conduct several ablation studies to confirm the effectiveness of our method.
\subsection{Datasets and Setup}
\subsubsection{Pretrain}We use 20 million unlabeled molecules to pretrain 3D-Mol. 
The unlabeled data is extracted from ZINC20\citep{irwin2020zinc20} and PubChem\cite{wang2009pubchem}, both of which are publicly accessible databases containing drug-like compounds. 
The raw data obtained from ZINC20 and PubChem is provided in SMILES format. 
To convert SMILES into molecular conformations for our pretraining stage, we utilize RDKit, a versatile Python cheminformatics package. RDKit enables the transformation of SMILES into structured molecular forms. 
We employ its ETKDG method, which generates realistic 3D conformations by integrating experimental torsion data with geometric principles.
To ensure consistency with prior research\citep{zhou2023uni,fang2022geometry}, we randomly select 90$\%$ of these samples for training purposes, while the remaining 10$\%$ were set aside for evaluation. 
For our model, we use the Adam optimizer with a learning rate of 1e-3. The batch size is set to 256 for pretraining and 32 for finetuning. 
The hidden size of all models is 256. 
The geometric embedding dimension K is 64, and the number of angle domains is 8. The hyperparameters $\lambda_{conf}$ and $\lambda_{fp}$ are both set to 0.5. 
The details of the pretraining environment and are in Appendix B.
\begin{table*}[htbp]
\centering
\caption{\textbf{Benchmarking the 3D-Mol and other pretraining methods.} We compare the performance on the 7 molecular property prediction tasks, marking the best results in bold and underlining the second best.}
\label{table_pt}
\resizebox{\textwidth}{!}{
\begin{tabular}{cccccccc}  
\toprule
 & \multicolumn{4}{c}{Classification (ROC-AUC \% higher is better ↑)} & \multicolumn{3}{c}{Regression (RMSE, lower is better ↓)} \\
\cmidrule(lr){2-5} \cmidrule(lr){6-8}
Datasets& {BACE} & {SIDER} & {Tox21} & {ToxCast} & {ESOL} & {FreeSolv} & {Lipophilicity} \\
\# Molecules & {1513} & {1427} & {7831} & {8597} & {1128} & {643} & {4200} \\
\# Tasks & {1} & {27} & {12} & {617} & {1} & {1} & {1} \\
\midrule
\addlinespace 
N-GramRF & $\rm 0.779_{0.015}$ & $\rm 0.668_{0.007}$ & $\rm 0.743_{0.004}$ & $-$ & $\rm 1.074_{0.107}$ & $\rm 2.688_{0.085}$ & $\rm 0.812_{0.028}$ \\[2pt]
N-GramXGB & $\rm 0.791_{0.013}$ & $\rm 0.655_{0.007}$ & $\rm 0.758_{0.009}$ & $-$ & $\rm 1.083_{0.107}$ & $\rm 5.061_{0.744}$ & $\rm 2.072_{0.030}$ \\[2pt]
PretrainGNN & $\rm 0.845_{0.007}$ & $\rm 0.627_{0.008}$ & $\rm 0.781_{0.006}$ & $\rm 0.657_{0.006}$ & $\rm 1.100_{0.006}$ & $\rm 2.764_{0.002}$ & $\rm 0.739_{0.003}$ \\[2pt]
\hline
& & & & & & & \\[-7pt]
3D Infomax & $\rm 0.797_{0.015}$ & $\rm 0.606_{0.008}$ & $\rm 0.644_{0.011}$ & $\rm 0.745_{0.007}$ & $\rm 0.894_{0.028}$ & $\rm 2.337_{0.107}$ & $\rm 0.695_{0.012}$ \\[2pt]
GraphMVP & $\rm 0.812_{0.009}$ & $\rm 0.639_{0.012}$ & $\rm 0.759_{0.005}$ & $\rm 0.631_{0.004}$ & $\rm 1.029_{0.033}$ & $-$ & $\rm 0.681_{0.010}$ \\[2pt]
$\rm GROVER_{base}$&$\rm 0.826_{0.007}$&$\rm 0.648_{0.006}$&$\rm 0.743_{0.001}$&$\rm 0.654_{0.004}$&$\rm 0.983_{0.090}$&$\rm 2.176_{0.052}$&$\rm 0.817_{0.008}$ \\[2pt]
$\rm GROVER_{large}$&$\rm 0.810_{0.014}$&$\rm 0.654_{0.001}$&$\rm 0.735_{0.001}$&$\rm 0.653_{0.005}$&$\rm 0.895_{0.017}$&$\rm 2.272_{0.051}$&$\rm 0.823_{0.010}$ \\[2pt]
$\rm MolCLR$&$\rm 0.824_{0.009}$&$\rm 0.589_{0.014}$&$\rm 0.750_{0.002}$&$-$&$\rm 1.271_{0.040}$&$\rm 2.594_{0.249}$&$\rm 0.691_{0.004}$ \\[2pt]
$\rm GEM$&$\rm 0.856_{0.011}$&$\rm \textbf{0.672}_{0.004}$&$\rm 0.781_{0.005}$&$\rm 0.692_{0.004}$&$\rm 0.798_{0.029}$&$\rm 1.877_{0.094}$&$\rm 0.660_{0.008}$ \\[2pt]
$\rm Uni$-$\rm Mol$ & $\rm \underline{0.857}_{0.005}$ & $\rm \underline{0.659}_{0.013}$ & $\rm \textbf{0.796}_{0.006}$ & $\rm  \underline{0.696}_{0.001}$ & $\rm\underline{0.788}_{0.029}$ & $\rm \underline{1.620}_{0.035}$ & $\rm \underline{0.603}_{0.010}$ \\[2pt]
\hline
& & & & & & & \\[-7pt]
$\rm 3D$-$\rm Mol$ & $\rm \textbf{0.872}_{0.004}$ & $\rm 0.658_{0.003}$ & $\rm \underline{0.792}_{0.003}$ & $\rm \textbf{0.701}_{0.003}$ & $\rm \textbf{0.782}_{0.008}$ & $\rm \textbf{1.617}_{0.050}$ & $\rm \textbf{0.600}_{0.015}$ \\[0pt]
\bottomrule
\end{tabular}
}
\end{table*}
\subsubsection{Finetune}We use 7 molecular property prediction datasets obtained from MoleculeNet to demonstrate the effectiveness of 3D-Mol. These datasets encompass a range of biophysics, physical chemistry and physiology. 
The details of the datasets are as followings:
\begin{itemize}
 \item \textbf{BACE}. The BACE dataset provides both quantitative (IC50) and qualitative (binary label) binding results for a set of inhibitors targeting human $\beta$-secretase 1 (BACE-1).
 \item \textbf{Tox21}. The Tox21 initiative aims to advance toxicology practices in the 21st century and has created a public database containing qualitative toxicity measurements for 12 biological targets, including nuclear receptors and stress response pathways.
 \item \textbf{Toxcast}. ToxCast, an initiative related to Tox21, offers a comprehensive collection of toxicology data obtained through in vitro high-throughput screening. It includes information from over 600 experiments and covers a large library of compounds.
 \item \textbf{SIDER}. The SIDER database is a compilation of marketed drugs and their associated adverse drug reactions (ADRs), categorized into 27 system organ classes.
 \item \textbf{ESOL}. The ESOL dataset is a smaller collection of water solubility data, specifically providing information on the log solubility in mols per liter for common organic small molecules.
 \item \textbf{FreeSolv}. The FreeSolv database offers experimental and calculated hydration-free energy values for small molecules dissolved in water.
 \item \textbf{Lipo}. Lipophilicity is a crucial characteristic of drug molecules that affects their membrane permeability and solubility. The Lipo dataset contains experimental data on the octanol/water distribution coefficient (logD at pH 7.4).
\end{itemize}
Extending from previous studies, we partition our datasets into training, validation, and test sets in an 80/10/10 ratio using scaffold splitting. 
This method groups molecules by their core structures, ensuring that each set features unique chemical scaffolds, in contrast to random splitting which allocates data indiscriminately of molecular similarity.
This approach rigorously tests the model on novel chemical entities, offering a more stringent evaluation of its generalization capabilities. 
We report the mean and standard deviation by the results of 3 random seeds.
The details of the finetuning settings and are in Appendix C.
\subsection{Metric}
In alignment with previous research, we employ the area under the receiver operating characteristic curve (ROC-AUC) as our evaluation metric for classification datasets.
ROC-AUC is a prevalent and reliable measure for gauging the effectiveness of binary classification models. 
For regression datasets, we apply root-mean-squared-error (RMSE) as our assessment metric, which is a standard for evaluating the accuracy of regression models in predicting continuous variables.
\subsection{Result}
\subsubsection{Overall performance}
To validate the efficacy of our method, we compare it with several baseline methods.
The baseline methods are as follows: 
N-Gram\citep{liu2019n} generated a graph representation by constructing node embeddings based on short walks. 
PretrainGNN\citep{huStrategiesPretrainingGraph2019} implemented several types of self-supervised learning tasks. 
3D Infomax\citep{stark20223d} maximized the mutual information between learned 3D summary vectors and the representations of a GNN. 
MolCLR\citep{wangMolecularcontrastivelearning2022} is a 2D-2D view contrastive learning model that involves atom masking, bond deletion, and subgraph removal. 
GraphMVP\citep{liu2021pre} used 2D-3D view contrastive learning approaches. 
GROVER\citep{NEURIPS2020_94aef384} focuses on node and graph level representation and corresponding pretraining tasks for each level. 
\begin{table*}[t]
\centering
\caption{\textbf{Benchmarking the 3D-Mol encoder and other non-pretraining methods.} We compare the performance on the 7 molecular property prediction tasks, marking the best results in bold and underlining the second best.}
\label{table_nopt}
\resizebox{\textwidth}{!}{
\begin{tabular}{cccccccc}  
\toprule
 & \multicolumn{4}{c}{Classification (ROC-AUC \% higher is better ↑)} & \multicolumn{3}{c}{Regression (RMSE, lower is better ↓)} \\
\cmidrule(lr){2-5} \cmidrule(lr){6-8}
Datasets & {BACE} & {SIDER} & {Tox21} & {ToxCast} & {ESOL} & {FreeSolv} & {Lipophilicity} \\
\# Molecules & {1513} & {1427} & {7831} & {8597} & {1128} & {643} & {4200} \\
\# Tasks & {1} & {27} & {12} & {617} & {1} & {1} & {1} \\
\midrule
\addlinespace 
$\rm DMPNN$&$\rm 0.809_{0.006}$&$\rm 0.570_{0.007}$&$\rm 0.759_{0.007}$&$\rm 0.655_{0.003}$&$\rm 1.050_{0.008}$&$\rm 2.082_{0.082}$&$\rm 0.683_{0.016}$ \\[2pt]
$\rm Attentive FP$&$\rm 0.784_{0.000}$&$\rm 0.606_{0.032}$&$\rm 0.761_{0.005}$&$\rm 0.637_{0.002}$&$\rm 0.877_{0.029}$&$\rm 2.073_{0.183}$&$\rm 0.721_{0.001}$ \\[2pt]
\hline
& & & & & & & \\[-7pt]
$\rm MGCN$&$\rm 0.734_{0.008}$&$\rm 0.587_{0.019}$&$\rm 0.741_{0.006}$&$-$&$-$&$-$&$-$ \\[3pt]
$\rm SGCN$&$-$&$\rm 0.559_{0.005}$&$\rm 0.766_{0.002}$&$\rm 0.657_{0.003}$&$\rm 1.629_{0.001}$&$\rm 2.363_{0.050}$&$\rm 1.021_{0.013}$ \\[2pt]
$\rm HMGNN$&$-$&$\rm \underline{0.615}_{0.005}$&$\rm 0.768_{0.002}$&$\rm 0.672_{0.001}$&$\rm 1.390_{0.073}$&$\rm 2.123_{0.179}$&$\rm 2.116_{0.473}$ \\[2pt]
$\rm DimeNet$&$-$&$\rm 0.612_{0.004}$&$\rm \underline{0.774}_{0.006}$&$\rm 0.637_{0.004}$&$\rm 0.878_{0.023}$&$\rm 2.094_{0.118}$&$\rm 0.727_{0.019}$ \\[2pt]
$\rm GEM$&$\underline{0.828}_{0.012}$&$\rm 0.606_{0.010}$&$\rm {0.773}_{0.007}$&$\rm \underline{0.675}_{0.005}$&$\rm \underline{0.832}_{0.010}$&$\rm \underline{1.857}_{0.071}$&$\rm \underline{0.666}_{0.015}$\\[2pt]
\hline
& & & & & & & \\[-7pt]
$\rm 3D$-$\rm Mol_{w.o\ pretrain}$&$\rm \textbf{0.839}_{0.005}$&$\rm \textbf{0.648}_{0.013}$&$\rm \textbf{0.790}_{0.004}$&$\rm \textbf{0.695}_{0.007}$&$\rm \textbf{0.807}_{0.027}$&$\rm \textbf{1.667}_{0.037}$&$\rm \textbf{0.620}_{0.004}$ \\[0pt]
\bottomrule
\end{tabular}
}
\end{table*}
GEM\citep{fang2022geometry} employed predictive geometry self-supervised learning schemes that leverage 3D molecular information. 
\begin{table*}[htbp]
\centering
\caption{\textbf{Ablation study.} We study the performance of 3D-Mol in four scenarios: 3D-Mol, 3D-Mol without pretraining, 3D-Mol without weight of contrastive learning, 3D-Mol without dihedral-angle graph, then mark the best results in bold and underline the second best.}
\label{table_ab}
\resizebox{\textwidth}{!}{
\begin{tabular}{cccccccc}  
\toprule
& \multicolumn{4}{c}{Classification (ROC-AUC \% higher is better ↑)} & \multicolumn{3}{c}{Regression (RMSE, lower is better ↓)} \\
\cmidrule(lr){2-5} \cmidrule(lr){6-8}
Datasets & {BACE} & {SIDER} & {Tox21} & {ToxCast} & {ESOL} & {FreeSolv} & {Lipophilicity} \\
\# Molecules & {1513} & {1427} & {7831} & {8597} & {1128} & {643} & {4200} \\
\# Tasks & {1} & {27} & {12} & {617} & {1} & {1} & {1} \\
\midrule
\addlinespace 
$\rm 3D$-$\rm Mol$ & $\rm \textbf{0.872}_{0.004}$ & $\rm \textbf{0.658}_{0.003}$ & $\rm \textbf{0.792}_{0.003}$ & $\rm \textbf{0.701}_{0.003}$ & $\rm \textbf{0.782}_{0.008}$ & $\rm \textbf{1.617}_{0.050}$ & $\rm \textbf{0.600}_{0.015}$ \\[3pt]
$\rm 3D$-$\rm Mol_{w.o\ pretrain}$&$\rm 0.839_{0.005}$&$\rm 0.648_{0.013}$&$\rm 0.790_{0.004}$&$\rm 0.695_{0.007}$&$\rm 0.807_{0.027}$&$\rm \underline{1.667}_{0.037}$&$\rm 0.620_{0.004}$ \\[3pt]
$\rm 3D$-$\rm Mol_{w.o. cl-weight}$&$\rm \underline{0.851}_{0.003}$&$\rm 0.645_{0.009}$&$\rm 0.786_{0.005}$&$\rm 0.696_{0.002}$&$\rm \underline{0.795}_{0.016}$&$\rm 1.705_{0.038}$&$\rm \underline{0.612}_{0.010}$ \\[3pt]
$\rm 3D$-$\rm Mol_{w.o. dihes-angle-graph}$&$\rm 0.844_{0.004}$&$\rm \underline{0.649}_{0.006}$&$\rm \underline{0.791}_{0.005}$&$\rm \underline{0.698}_{0.006}$&$\rm 0.812_{0.015}$&$\rm 1.782_{0.007}$&$\rm \underline{0.612}_{0.005}$ \\[0pt]
\bottomrule
\end{tabular}
}
\end{table*}
Uni-Mol\citep{zhou2023uni} enlarged the application scope and representation ability of molecular representation learning by using transformer. 
Table \ref{table_pt} presents compelling evidence that methods based on 3D information surpass those based on 2D in molecular modeling, as evidenced by the improved outcomes across multiple datasets. 
Besides, compared to current 3D-based approaches, our method achieves the best results in 5 datasets and the second-best in Tox21, highlighting its exceptional performance. 
Notably, our method exhibits a remarkable lead in BACE dataset. 
This not only affirms the value of considering spatial configurations in predictive models but also indicates our method's potential in utilizing this information for high-precision molecular property predictions. 
Moreover, our method's dominance extends across all datasets, except for some toxicity datasets. In these cases, its focus on geometric rather than substructural information, which is crucial for toxicity prediction, suggests an avenue for further refinement.
\subsubsection{Encoder performance}
To validate the efficacy of 3D-Mol encoder, we compare it with several baseline molecular encoder that do not employ pretraining.
The baseline molecular encoders are as follows: 
DMPNN\citep{yang2019analyzing} employed a message passing scheme for molecular property prediction. 
AttentiveFP\citep{xiong2019pushing} is an attention-based GNN that incorporates graph-level information. 
MGCN\citep{lu2019molecular} designed a hierarchical GNN to directly extract features from conformation and spatial information, followed by multilevel interactions. 
HMGNN\citep{shui2020heterogeneous} leverages global molecular representations through an attention mechanism.
SGCN\citep{danel2020spatial} applies different weights according to atomic distances during the message passing process. 
DimeNet\citep{gasteiger2020directional} proposed directional message passing to fully utilize directional information within molecule. 
GEM\citep{fang2022geometry} employed message passing strategy to extract 3D molecular information. 
As the results shown in Table \ref{table_nopt}, 3D-Mol encoder significantly outperforms all the baselines on both types of tasks and improves the performance over the best baselines with 2$\%$ and 11$\%$ for classification and regression tasks, respectively, since 3D-Mol incorporates geometrical parameters.
\subsubsection{Ablation study}
To validate the efficacy of our pretraining task, we study the performance of the 3D-Mol encoder in three scenarios: 3D-Mol, 3D-Mol without pretraining, and 3D-Mol without the dihedral-angle graph. 
The results are shown in Table \ref{table_ab}.
Compared with the 3D-Mol and 3D-Mol without pretraining, the former performs better in all datasets, demonstrating that our pretraining method can improve encoder performance.
Compared to the version without contrastive learning weights from fingerprints and 3D descriptors, 3D-Mol demonstrates superior performance across all datasets. This improvement shows that using weights from fingerprints and 3D descriptors effectively optimizes our contrastive loss, enhancing encoder performance.
Similarly, Compared with the 3D-Mol and 3D-Mol without $G_{d-a}$, the former also shows better performance in all datasets, indicating that the dihedral-angle graph contributes to improved encoder performance.
In general, our pretraining and modeling methods enhance the 3D-Mol encoder performance, as the model can more effectively learn the 3D structural information of molecule.
\begin{figure}[t!]
\centering
    \includegraphics[width=1.0\linewidth]{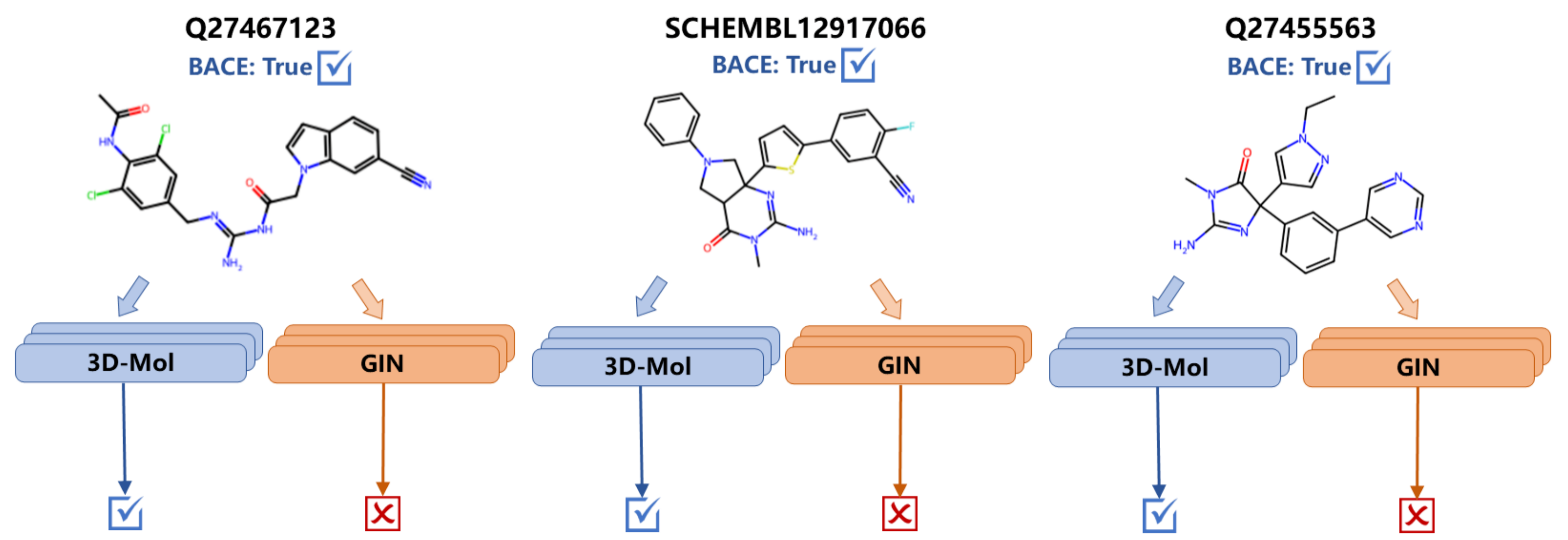}
    \caption{\textbf{Case study of 3D information enhance using 3D-Mol for the BACE task.} This figure illustrates the prediction results for three molecules identified as BACE inhibitors, showcasing the efficacy of 3D-Mol, which utilizes three-dimensional molecular information, versus GIN, which relies on two-dimensional data. Each molecule is displayed with its respective name (Q27467123, SCHEMBL12917066, Q27455563), and the results indicate that 3D-Mol accurately predicts BACE inhibition across all cases, while GIN fails to do so.}
    \label{fig:cs}
\end{figure}
\begin{figure}[t!]
\centering
    \includegraphics[width=1.0\linewidth]{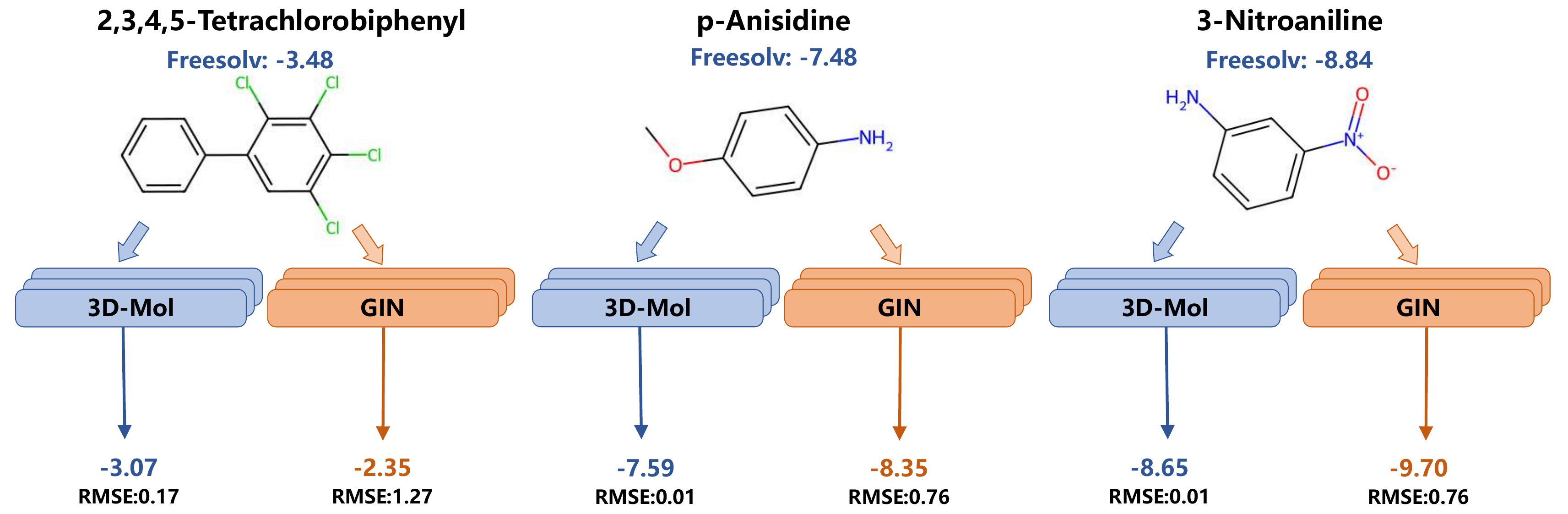}
    \caption{\textbf{Case study of 3D information enhance using 3D-Mol for the Freesolv task.} This figure illustrates the prediction results for hydration free energy of three molecules, showcasing the efficacy of 3D-Mol, which utilizes three-dimensional molecular information, versus GIN, which relies on two-dimensional data. Each molecule is displayed with its respective name (2,3,4,5-Tetrachlorobiphenyl, p-Anisidine, 3-Nitroaniline), and the results indicate that 3D-Mol accurately predicts hydration free energy across all cases, while GIN fails to do so.}
    \label{fig:cs2}
\end{figure}
\subsubsection{Case Study}
In this case study, we explore the predictive capabilities of 3D-Mol, which utilizes three-dimensional molecular data, compared to GIN, which relies solely on two-dimensional information, focusing on their efficacy in identifying inhibitors of the $\beta$-secretase 1 (BACE) enzyme, a crucial target for Alzheimer's disease treatment. 
We specifically analyze three molecules: Q27467123, which features a chlorinated aromatic ring that may form key halogen bonds within the BACE active site; SCHEMBL12917066, with a fluorinated aromatic ring that enhances molecular stacking interactions; and Q27455563, a molecule with a bicyclic structure potentially forming multiple hydrogen bonds. 
As shown in Figure \ref{fig:cs}, 3D-Mol's approach, which integrates these 3D conformations and potential interactions such as hydrogen bonding, hydrophobic contacts, and precise geometric fitting into the BACE active site, leads to accurate predictions of their inhibitory activities. 
In contrast, GIN, unable to account for such intricate 3D-dependent interactions, fails to recognize these molecules as potential inhibitors, demonstrating limitations in capturing the necessary depth and spatial relationships that are crucial for binding. 
This case study highlights the importance of incorporating 3D spatial awareness in computational models for drug discovery, particularly when targeting enzymes like BACE, where the precise alignment and interaction of molecules within a complex three-dimensional space significantly influence their therapeutic efficacy.\\
\indent Having validated our 3D-Mol framework with the BACE dataset, we extend its application to the Freesolv dataset, which is instrumental for predicting the hydration free energy of small molecules in water—a crucial determinant for solubility in drug discovery and molecular design. We present a case study of three molecules with distinctive 3D structures that influence their solvation behaviors: 2,3,4,5-Tetrachlorobiphenyl, noted for its hydrophobicity due to chlorine substitution; p-Anisidine, which can engage in hydrogen bonding; and 3-Nitroaniline, where the nitro group impacts solvation through electron withdrawal.
Figure \ref{fig:cs2} displays a comparative analysis of the 3D-Mol and GIN model predictions for these molecules. The Freesolv values and RMSE scores indicate 3D-Mol's superior ability to capture the impact of molecular geometry on solvation. This case study highlights the necessity of 3D information for precise prediction of solvation-related molecular properties, reinforcing the utility of the 3D-Mol framework in computational chemistry.

\section{Conclusion and Discussion}
\label{sec:con}
3D-Mol, our novel framework introduced in this paper, significantly advances molecular property prediction by leveraging a unique hierarchical graph-based embedding and a contrastive learning component. This approach allows for a comprehensive capture of 3D molecular structures, setting a new standard in the field. Demonstrating superior performance over multiple benchmark models, 3D-Mol holds immense potential in revolutionizing AI-assisted drug discovery and molecular design.

3D-Mol's encoder distinctly outperforms all baselines, with performance improvements of 2\% in classification and 11\% in regression tasks compared to traditional and non-pretraining methods. These advancements are pivotal in drug discovery, as they promise to expedite the development of new treatments through more accurate molecular predictions. The primary challenge faced by 3D-Mol is the time-intensive generation of 3D conformations and the encoding of hierarchical graphs. Future work will focus on optimizing these processes to enhance the model's efficiency and practical scalability, further solidifying its role in advancing computational chemistry and pharmaceutical development.

\section*{Acknowledgment}
The research was supported by the Peng Cheng Laboratory and by Peng Cheng Laboratory Cloud-Brain.

\bibliographystyle{unsrt} 
\bibliography{arxiv.bib}

\begin{appendices}


\section{3D Conformation Descriptor and Fingerprint}
\subsection{Fingerprint}
In our study, we integrate molecular fingerprints, particularly Morgan fingerprints, to calculate weights for negative pairs in our model. 
These fingerprints, which provide a compact numerical representation of molecular structures, are crucial for computational chemistry tasks. 
The Morgan fingerprint method iteratively updates each atom's representation based on its chemical surroundings, resulting in a detailed binary vector of the molecule. 
By evaluating the similarity between Morgan fingerprints, we derive a precise weighting mechanism for negative pairs, enhancing our model's ability to detect and differentiate molecular structures. 
This methodology not only improves our model's accuracy in molecular interaction analysis but also adds to its overall predictive capabilities.
\subsection{3D Conformation Descriptor}
Molecular 3D conformation descriptors are computational tools used to represent the three-dimensional arrangement of atoms within a molecule, capturing critical aspects of its spatial geometry. 
These descriptors are crucial in understanding how molecular shape influences chemical and biological properties, and they play a significant role in fields like drug design and materials science.
The 3D-Morse descriptor, specifically, is a type of 3D molecular descriptor that quantifies the molecular structure using electron diffraction patterns, offering a unique approach to encapsulating the spatial distribution of atoms. 
It provides a detailed and nuanced representation of molecular conformation, making it highly valuable in computational chemistry and cheminformatics.
In our research, we employ 3D-Morse descriptors to measure the similarity of molecular 3D conformations, enabling us to compare and analyze molecular structures effectively and identify potential similarities in their biological or chemical behavior. 
This application of 3D-Morse descriptors is instrumental in fields such as drug discovery, where understanding molecular similarities can lead to the identification of new therapeutic compounds or the prediction of their activities.
\section{The contribution of pretraining method}
\begin{table*}[htbp]
\centering
\caption{\textbf{The contribution of pretraining method.} We study the performance of 3D-Mol in three scenarios: contrastive learning only, supervised pretraining only, complete pretraining method, then mark the best results in bold and underline the second best.}
\label{table_ab_2}
\resizebox{\textwidth}{!}{
\begin{tabular}{cccccccc}  
\toprule
& \multicolumn{4}{c}{Classification (ROC-AUC \% higher is better ↑)} & \multicolumn{3}{c}{Regression (RMSE, lower is better ↓)} \\
\cmidrule(lr){2-5} \cmidrule(lr){6-8}
Datasets & {BACE} & {SIDER} & {Tox21} & {ToxCast} & {ESOL} & {FreeSolv} & {Lipophilicity} \\
\# Molecules & {1513} & {1427} & {7831} & {8597} & {1128} & {643} & {4200} \\
\# Tasks & {1} & {27} & {12} & {617} & {1} & {1} & {1} \\
\midrule
\addlinespace
$\rm Contrastive-Learning-Only $&$\rm 0.847_{0.002}$&$\rm \underline{0.652}_{0.012}$&$\rm 0.791_{0.008}$&$\rm 0.693_{0.003}$&$\rm 0.802_{0.036}$&$\rm 1.682_{0.86}$&$\rm 0.616_{0.023}$ \\[3pt]
$\rm Supervised-Pretraining-Only $&$\rm \underline{0.862}_{0.006}$&$\rm 0.647_{0.007}$&$\rm \textbf{0.796}_{0.002}$&$\rm \underline{0.697}_{0.003}$&$\rm \underline{0.795}_{0.022}$&$\rm \underline{1.664}_{0.070}$&$\rm \underline{0.613}_{0.004}$ \\[3pt]
$\rm Complete-Pretraining-Method $&$\rm \textbf{0.872}_{0.004}$ & $\rm \textbf{0.658}_{0.003}$ & $\rm \underline{0.792}_{0.003}$ & $\rm \textbf{0.701}_{0.003}$ & $\rm \textbf{0.782}_{0.008}$ & $\rm \textbf{1.617}_{0.050}$ & $\rm \textbf{0.600}_{0.015}$ \\[0pt]
\bottomrule
\end{tabular}
}
\end{table*}
In this section, we discuss the contributions of contrastive learning and supervised pretraining methods to our pretraining approach. 
We pretrained our model using three approaches: contrastive Learning only, supervised pretraining only, and complete pretraining method. We compared their performance on 7 benchmark datasets. 
As the Table \ref{table_ab_2} shown, the contributions of both contrastive learning and supervised pretraining were less significant than the complete method. These findings emphasize that while both contrastive learning and supervised pretraining contribute positively to the model's performance, their combination is crucial for achieving optimal results.
\section{Finetuning Details}
During finetuning for each downstream task, we randomly search the hyper-parameters to find the best performing setting on the validation set and report the results on the test set. 
Table \ref{tabel_hp} lists the combinations of different hyper-parameters. 
\begin{table*}[htbp]
    \centering
    \renewcommand\arraystretch{1}
    \caption{\textbf{hyper-parameter setting}}
    \label{tabel_hp}
    \resizebox{0.75\textwidth}{!}{
    \begin{tabular}{c c c}  
        \toprule[2pt]
    	& & \\[-6pt]
    	Name& Description& Range \\  
    	\midrule[1pt]
    	   & & \\[-6pt] 
    	   $lr_{MLP}$&Initial learning rate for MLP head&$\{0.004, 0.001, 0.0004\}$ \\
            \vspace{1mm}
    	   & & \\[-6pt]  
    	   $lr_{ENC}$&Initial learning rate for the pre-trained encoder&$\{0.001, 0.0004, 0.0001\}$ \\
            \vspace{1mm}
    	   & & \\[-6pt] 
    	   $Epoch$&The number of epoch in finetuning stage& $\{60, 80, 100\}$ \\
            \vspace{1mm}
    	   & & \\[-6pt] 
    	   $num_{layer}$&Number of hidden layers in MLP&$\{2, 3\}$ \\
            \vspace{1mm}
    	   & & \\[-6pt] 
    	   $Dropout$&Dropout ratio for the model&$\{0, 0.1, 0.2, 0.5\}$ \\
            \vspace{1mm}
    	   & & \\[-6pt]   
    	   $Hidden_{size}$&Size of hidden layers in MLP&$\{32, 64, 128, 256\}$ \\
        \toprule[2pt]
    \end{tabular}
    }
\end{table*}
\section{Environment}
\textbf{CPU}:\\
$\bullet$ Architect: $X86 $ $ 64$ \\
$\bullet$ Number of CPUs: 96 \\
$\bullet$ Model: Intel(R) Xeon(R) Platinum 8268 CPU @ 2.90GHz \\
\\
\textbf{GPU}: \\
$\bullet$ Type: Tesla V100-SXM2-32GB \\
$\bullet$ Count: 8 \\
$\bullet$ Driver Version: 450.80.02 \\
$\bullet$ CUDA Version: 11.7 \\
\\
\textbf{Software Environment}: \\
$\bullet$ Operating System: Ubuntu 20.04.6 LTS \\
$\bullet$ Python Version: 3.10.9 \\
$\bullet$ Paddle Version: 2.4.2 \\
$\bullet$ PGL Version: 2.2.5 \\
$\bullet$ RDKit Version: 2023.3.2 \\



\end{appendices}

\end{document}